# On Quantum Bayesianism

Moses Fayngold
*Department of Physics, New Jersey Institute of Technology, Newark, NJ 07102*

The lately developed part of Quantum Bayesianism named QBism has been proclaimed by its authors a powerful interpretation of Quantum Physics. This article presents analysis of some aspects of QBism. The considered examples show inconsistencies in some basic statements of the discussed interpretation. In particular, the main quantum mechanical conundrum of measurement and the observer is, contrary to the claims, not resolved within the framework of QBism. The conclusion is made that the basic tenets of QBism as applied in Physics are unsubstantiated.

Kea words:   QBism, probability, degree of belief, quantum non-locality

> The subject matter of the theory is not the world or us
> but us-within-the-world, the interface between the two.
>                                             Christopher Fuchs

## 1. Introduction

  I will agree 100% with this epigraph (cited from [1]) if I can paraphrase it as "The subject matter of *all science* is *the world and us-within-it*", emphasizing that all human scientific endeavors have been attempts to solve the Mystery of Being. Both, *us and the world*, are equally mysterious and may be linked much more intimately than we know today.
  Quantum Mechanics (QM) has opened new venues in our study of the world by discovering its intrinsically probabilistic nature, indeterminacy, and the role of measurement and human observer. This might also be a crucial step towards grasping a mystery of human consciousness.
 "Quantum Bayesianism (QBism) [2, 3] or "Quantum Bettabilitarianism" in its latest version [4] is known as an attempt to resolve some controversial topics in QM by promoting the role of an "*agent*" (human observer).
 The presented article discusses some basic statements of QBism as described in [1-3] and in the most recent publications [4-8]. I will analyze those aspects of QBism that seem to me debatable.
   The quotes from the discussed or referred sources are written in italics and in quotation marks. My comments are mostly in the regular font, but in some cases which I want to emphasize, I also use italics.

## 2. Observer, an object, and object's state

{1} *"The basic statement of the theory (conventional quantum mechanics (QM)) - the one we have all learned from our textbooks - seems to rely on terms our intuitions balk at as having any place in a fundamental description of reality. The notions of "observer" and "measurement" are taken as primitive, the very starting point of the theory. This is an unsettling situation! Shouldn't physics be talking about <u>what is</u> before it starts talking about what will be seen and who will see it?"* ([2], p. 1)

Compare this with

{2} "*According to QBism, quantum mechanics is a tool anyone can use to evaluate, on the basis of one's past experience, one's probabilistic expectations for one's subsequent experience. Unlike Copenhagen, QBism explicitly takes the "subjective" or "judgmental" or "personalist" view of probability… : probabilities are assigned to an event by an agent* [1] *and are particular to that agent. The agent's probability assignments express her own personal degrees of belief about the event*." ([3], p. 1, 2)

These two quotations contradict each other. The first promotes "*what is*" to the status of primary concept, whereas the second appears to grant this status to the "*agent*" as "*one who acts*", that is, to *observer*.

The question posed in {1} calls for a counter-question: How could anyone talk about *what is* before learning about *its existence* through *observing it* or *its manifestations*? A conventional view in Physics is that an object or system is determined by its properties. What *is* known to us out there had been first perceived by our senses either directly or indirectly through its manifestations. And the next step is verifying the observations by comparing them with those made by other observers under similar conditions. An entity that cannot in principle be observed directly or indirectly is as good as non-existing.

So it is not immediately obvious what should be more "*primitive*" – "*what is*" or "*observer and measurement*". But even though the *observation* is indeed far more important in QM than in classical physics, it is not taken as something superior over "*what is*". Conventional QM emphasizes the objectivity of observed systems even when some of their characteristics such as quantum indeterminacy surpass all previously learned concepts. This unusual nature of QM may seem unsettling to purely classical mind, but we must accept the verified experimental evidence even it goes beyond our classical intuition. So both – *the observer* and *what is observed* – are equally important.

Actually, the notions of "observer" and "measurement" had been promoted long before the emergence of QM. The observer-dependence of velocity or kinetic energy has been known from Galileo's time. Yet each characteristic is still objective because its value measured in any given reference frame (RF) is the same for all observers in this frame.

The Bayesian's claim that quantum states are merely observer's expectations ("*degrees of belief*") rather than objective characteristics of real systems undermines their other claim promoting the status of "*what is*".

One could argue that there are no contradictions here since {1} talks about an *object* ("*what is*") whereas {2} – about an *event*. But an event is an abrupt change *in a system*, and making it dependent on agent's degree of belief undermines the system's objectivity. And more important, degree of belief itself, as mentioned above, originates from the previous measurements.

---

[1] "*Agent*" *in the sense of one who acts* (*and not in the sense of one who represents another*).

The fundamental role of observation and measurement is a specific feature of QM. A QM observation usually creates a new reality by radically changing the state of the observed/measured object, rather than just revealing to us already pre-existing state. This happens because of the QM indeterminacy and generally irreducible interactions between the measured object and the environment (including experimental setup). As a result, we may have a series of different possible measurement outcomes in a pure ensemble of objects. The probability of each outcome is an objective characteristic of the preexisting state. The so-called interaction-free measurements studied in QM are also probabilistic and, in addition, nonlocal [9-11]. Their existence does not affect our arguments.

*Probabilistic nature* of the world is intimately connected with Special Relativity (SR). The quantum world can be relativistic only by being intrinsically probabilistic [12]. If a classical measurement just verifies what had already existed before the observation, QM measurement generally actualizes one out of the set of potentialities which had existed in a state of superposition. In this respect, the measurement abruptly changes the reality (produces new information in QBism's language [2]). We might call such process "projective creation" (paraphrasing von Neumann's projective measurements [13]). The most puzzling feature here is that QM, while exactly predicting probability of any possible outcome, does not describe any mechanism of its actual selection. It even postulates that it is meaningless to ask about such mechanism since the latter is not manifest in any known observations. The theory answers all relevant questions about the observed reality without references to any "mechanisms". This has produced in many an impression of crucial and may be even superior role of observation and measurement. It is best encapsulated in John Bell's question [14]:

{3} "*What exactly qualifies some physical systems to play the role of 'measurer'? Was the wave function of the world waiting to jump for thousands of millions of years until a single-celled living creature appeared? Or did it have to wait a little longer, for some better qualified system . . . with a PhD?*"

In other words, what is this elusive something that interrupts the continuous deterministic evolution of a state vector under the given Hamiltonian and causes its discontinuous jump into one of the corresponding eigenstates? Such a jump is usually called the collapse of a wave function. This widely used term is largely misleading because it totally ignores the opposite side of the coin: what is the collapse with respect to a measured observable may be an *explosion* with respect to the complementary observable [15]. This is probably one of the most exotic features of QM, all the more so that the moment of the "jump" and its outcome can be predicted only probabilistically. Even some creators of quantum theory – Einstein, de Broglie, Schrodinger, – found it hard to come to terms with these new features. The situation provoked natural reactions including Bayesian claim like this:

{4} "*…quantum states are not something out there, in the external world, but instead are expressions of information. Before there were people using quantum theory as a* *branch of physics, before they were calculating neutron-capture cross-sections for uranium and working on all the other practical problems the theory suggests, there were no quantum states. The world may be full of stuff and things of all kinds, but among all the stuff and all the things, there is no unique, observer-independent, quantum-state kind of stuff.*" ([1, 3], pp. 2, 5, respectively)

In other words, quantum states (and thereby the resulting classical states) are only mental constructs in our minds. If so, can a QBist convincingly explain what were they all invented for? Had our planet existed in a state favorable for humans before humans or had there been just some undefined stuff that formed the known shape only after the appearance of humans?

A similar Bayesian statement with an emphasis on the *quantum* states is exemplified by the following quote [16]:

{5} "*A quantum-mechanical state being a summary of the observers' information about an individual physical system changes both by dynamical laws, and whenever the observer acquires new information about the system through the process of measurement. The existence of two laws for the evolution of the state vector becomes problematical only if it is believed that the state vector is an objective property of the system. If, however, the state of a system is defined as a list of* [*experimental*] *propositions together with their* [*probabilities of occurrence*], *it is not surprising that after a measurement the state must be changed to be in accord with* [*any*] *new information. The "reduction of the wave packet" does take place in the consciousness of the observer, not because of any unique physical process which takes place there, but only because the state is a construct of the observer and not an objective property of the physical system.*" ([2, 4], p. 2, 5)

The beginning of this statement "...*the observer acquires new information about the system through <u>the process of measurement</u>*" contradicts its end stating that "*The "reduction of the wave packet" takes place… only because the state is a construct of the observer*". The latter also contradicts the known fact that "…*reduction of the wave packet*" occurs because of a specific physical <u>process</u> – e.g., packet's interaction with a high-energy laser pulse crossing the same region [17, 18]. Another evidence of objectivity of reduction is firing of one of the detectors, which signals localization of the initial probability cloud of a studied particle. The reduction (considered in some details in [12]) happens instantaneously by our everyday standards and is not described by any wave equation. And yet QM says quite correctly that the abrupt change of probabilities merely reflects equally abrupt change of the measured state. There is no other option, simply because one unchanging state (or worse, no state at all) cannot be characterized by two different sets of probabilities for the same observable. And if, according to extreme assertion of QBism, there are no states at all, then probabilities *of what* are formed and changed in the observer's mind?

A quantum state $|p\rangle$ with definite value of observable *p* is a superposition of eigenstates of the complementary observable *q*,

$$|p\rangle = \int c_p(q)|q\rangle dq \;, \tag{1}$$

and an exact value $q = q'$ may appear only with probability $|c_p(q')|^2 \, dq$ in a subsequent $q$-measurement. But this does not undermine the objectivity of the resulting state. Generally, a state $|\Psi\rangle$ contains *exact* information about probabilities, variances, and expectation values of any relevant observable in this state. Just because the concept of a state turned out to have an additional dimension embracing QM indeterminacy does not make it unreal.

Of course, these comments are based on the conventional definition of a state. A state is determined by a set of characteristics of an object. A container with gas is defined by its mass, chemical composition, temperature, etc. Changing at least one of these characteristics will change the state, sometimes beyond recognition – e.g., it may become a cloud of plasma with container evaporated under raised temperature. One can reasonably ask if this is the same object as before. Many (myself included) will say that an electron-positron pair is a different object than a set of $\gamma$-photons emerging after its annihilation.

In any case, *it is a state that defines the object*. Remove the state from a photon – and what remains is the word "photon" with no real photon as such. The same is true in the classical limit of QM. Remove from a boulder its state – mass, shape, size, rigidity, etc. - and then tell me "*what is*" there, if there remains anything at all? The Bayesian answer "*stuff*" does not tell much to an inquiring mind.

All this can be summarized by a brief statement by Zurek: "*In quantum physics, reality can be attributed to the measured states*" [18].

One could object that this statement mentions only *measured* states. But any measurement, no matter how sophisticated, imitates a possible natural process. The Stern-Gerlach experiment in a spin measurement naturally occurs when a corresponding atom passes through an inhomogeneous magnetic field. The electron position measurement naturally occurs when an electron gets absorbed, say, by a positive ion. The only difference is that the outcomes of these natural measurements are usually not amplified to the macroscopically observable events to be recorded by humans. But they are recorded by Nature due to their imprints on evolution on the microscopic level.

In QM, an isolated object may show different faces depending on observer's actions. A quantum wave packet may show its particle aspect in position measurement or its wave aspect in momentum measurement. Either of these faces is real upon exposure, albeit brought to full life only by an appropriate measurement at the cost of "dissolving" the other face – they cannot coexist in full together. This "versatility" of appearances reflects richness of reality by far exceeding our classical intuition, and does illustrate a much more active role of an observer. To a certain degree, we can only see what we choose to observe. But this is only possible because various faces already preexist as potentialities, and each of them can emerge in full in the respective experimental setup.

In a most counterintuitive phenomenon of entanglement, the states of two objects are correlated so that neither state is completely defined independently from the other. Such inseparability of states appears to contradict our initial argument connecting an object with its state. But there is no contradiction here because an *incompletely defined* individual state does not amount to the *total absence* of a state. The state still exists but some of its characteristics necessary for *complete* description are contingent on those of the other object. The object is correlated with its counterpart in such a way that a certain measurement outcome for one instantaneously affects the other. This instantaneous effect may seem to be a sign of superluminal signaling, which would contradict SR. This, in turn, motivated efforts to reinterpret QM in such a way as to avoid the indicated "contradiction". But all such attempts, apart from being fruitless, were not justified to begin with, because quantum collapse of the system to completely defined individual states (disentanglement) does not contradict SR [12, 15] (some specific examples will be mentioned in the end of Sec. 2 and in Sec. 3).

{6} "*A proponent of the ontic view might argue that the phenomena in question are not mysterious if one abandons certain preconceived notions about physical reality. The challenge we offer to such a person is to present a few simple physical principles by the light of which all of these phenomena become conceptually intuitive* (*and not merely mathematical consequences of the formalism*) *within a framework wherein the quantum state is an ontic state. Our impression is that this challenge cannot be met. By contrast, a single information-theoretic principle, which imposes a constraint on the amount of knowledge one can have about any system, is sufficient to derive all of these phenomena in the context of a simple toy theory . . . "*  ([2, 4, 19], p. 3, 6)

The offered challenge cannot, indeed, be met for the reason already mentioned before: the concepts discussed are not (yet) intuitively clear – simply because the aspects of Nature they describe are not intuitively clear within the realm of our previous experience. The demand of *intuitively clear* formulation of QM is equivalent to requirement to describe a rainbow in terms of darkness familiar to one grown up in an underground cave.

The innately probabilistic features of reality unveil a deeper truth about Nature, by far surpassing a straightforward single-valued classical determinism. The "fuzziness" of position and/or momentum of a particle in a state $|\Psi\rangle$ is an objective characteristic of this state rather than observer's uncertainty about it. In this respect the misleading term "uncertainty principle" widely used in the English-speaking part of the Physics community turned out to have disastrous consequences implying that our knowledge of an object is restricted due to some hidden flaws in the communication between the object and observer. Such flaws, according to this view, keep some variables unknown to us. Hence the term "hidden variables" [20] implying that it is just the lack of information accessible to us that leads to QM uncertainty. In fact, $\Delta x$ and $\Delta p_x$ are *real indeterminacies* in position and momentum of an object rather than observer's uncertainty about them, and therefore the truly appropriate term describing them would be the "indeterminacy principle". The German "unbestimmheit" in the original formulation of the principle means just that – indeterminacy. The term "uncertainty" in its English formulation is just a sloppy translation.

Even the most counterintuitive predictions based on QM indeterminacy (e.g., Bell's theorem associated with quantum non-locality [20, 21]) have passed all experimental tests [22-25] with flying colors.

The assertion "*Quantum States Do Not Exist*" ([2], Sec. 2) promotes the previously quoted Bayesian concept of "*stateless stuff*" to a universal principle. What about the Bayesians themselves? Are they unique personalities with infinitely reach characteristics or just "*stuff*"?

Now we turn to a Bayesian view of probability as subjective expectations [26, 27], illustrated by the following dialogue:

{7} " ***Pre-Bayesian:*** *Ridiculous, probabilities are without doubt objective. They can be seen in the relative frequencies they cause.*
***Bayesian:*** *So if P = 0.75 for some event, after 1000 trials we'll see exactly 750 such events?*
***Pre-Bayesian:*** *You might, but most likely you won't see that exactly. You're just likely to see something close to it.*
***Bayesian:*** *Likely? Close? How do you define or quantify these things without making reference to your degrees of belief for what will happen?*
***Pre-Bayesian:*** *Well, in any case, in the infinite limit the correct frequency will definitely occur.*
***Bayesian:*** *How would I know? Are you saying that in one billion trials I could not possibly see an "incorrect" frequency? In one trillion?*
***Pre-Bayesian:*** *OK, you can in principle see an incorrect frequency, but it'd be ever less likely!*
***Bayesian:*** *Tell me once again, what does 'likely' mean?*" ([2], pp. 4, 5)

There are at least three objections to this kind of Bayesian's logic.

1) The "*likelihood*" sneered at by the Bayesian *is* objectively quantified, as correctly stated by Pre-Bayesian. The experimentally observed deviations from rigorously defined probability are also quantifiable (see, e.g. the Poisson distribution [28]). The fact that their rigorous definition involves a limiting procedure does not undermine their objectivity, as seen from the next example.

2) According to Bayesian, probability is not an objective characteristic because it is restricted in practice only to finite sets of trials. But the latter is true for any physical law! The Lorentz force law

$$\mathbf{F} = q\,\mathbf{v} \times \mathbf{B} \qquad (2)$$

for a charge *q* moving with velocity **v** in magnetic field **B** had been confirmed beyond any doubts. But if we plug into (2) the experimentally measured **F**, **v**, and **B**, we will practically never get the exact equality. It would become exact, just as probability $\mathcal{P}$ in the quoted dialog, only after the elimination of systematic errors and only after the averaging procedure in an infinite set of trials. Since the latter is never attainable, the Bayesian logic leads to conclusion that the Lorentz force law is not an objective reality. And by induction, the same logic denies objectivity of all laws of Nature.

It is true that physical equations are only simplified mathematical models of reality. Equations of planetary motions usually ignore the effects of SR. Equations of SR ignore the curvature of space-time, and so on. But all this does not undermine the objectivity of laws of Nature.

3) According to Bayesians, two different agents may give different assignments for probability of the same event, e.g., for finding a particle in a given energy state in a gas container at a fixed temperature. Each agent's assignment will depend on his/her expertise and many other factors. If probability is merely the subjective "*degree of belief*", then whose assignment will have priority?

The Bayesian may object that argument 3) is not warranted since the "agents" must be certified experts and they derive one common assignment from experimental results. But such objection would automatically transfer the Bayesian to the camp of conventional QM in which the Supreme Judge is a verifiable experiment.

The arguments 1) – 3) can be summarized in the following parody on the above-quoted conversation between the two sides. In this parody, I just swap their roles and terms "probability" and "*degree of belief*".

" **Bayesian**: Ridiculous, degrees of belief are without doubt most vital. They can be seen in the relative frequencies they cause.
**Pre-Bayesian**: So if you assign $D = 0.75$ to your degree of belief for some event, then after 1000 trials we'll see exactly 750 such events?
**Bayesian**: You might, but most likely you won't see that exactly. You're just likely to see something close to it.
**Pre-Bayesian**: Likely? Close? Is it another degree of your belief? How do you define or quantify these things without making reference to probabilities for what will happen?
**Bayesian**: Well, in any case, in the infinite limit the correct frequency will definitely occur.
**Pre-Bayesian**: How would I know? Are you saying that in one billion trials I could not possibly see an "incorrect" frequency? In one trillion?
**Bayesian**: OK, you can in principle see an incorrect frequency, but it'd be ever less likely!
**Pre-Bayesian**: Tell me once again, what does 'likely' mean?"

Identifying *probability* with subjective "*…degree of belief*" is not a scientific statement. A belief as such cannot be quantified, whereas probability has an exact quantitative definition [29-32]. The fact that a set of trials gives only its approximate value is of the same nature as an approximate value of the speed of light resulting from a finite set of measurements. And yet it is the physical measurements that determine the speed of light or gravitational constant or neutron's lifetime with ever-increasing accuracy.

The considered statements of QBism clash with QM and sometimes conflict even with each other like in examples {1}, {2}. Here is another example:

{8} "*Quantum states, through the Born Rule, can be used to calculate probabilities.*" ([2], p. 5)

This is true in conventional QM which assumes the objectivity of quantum states. But according to the Bayesian (statement {4} and Sec. 2 in [2]), quantum states do not exist! How can we calculate anything reliable from something not objectively existing? And whatever that calculated thing might be, what is it good for?

{9} *"Conversely, if one assigns probabilities for the outcomes of a well-selected set of measurements, then this is mathematically equivalent to making the quantum-state assignment itself. The two kinds of assignments determine each other uniquely."*
([2], p. 5)

This is, at best, ambiguous. What kind of a set is "*well-selected*"? And what are the advantages of this proposal? Consider a superposition:

$$|\Psi(t)\rangle = \sum_j \tilde{a}_j |q_j\rangle, \quad \tilde{a}_j \equiv a_j e^{i(\alpha_j - \omega_j t)} \qquad (3)$$

where $q_j$, $|q_j\rangle$ are, respectively, the $j$-th eigenvalue and eigenstate of observable $q$. Knowing $|\Psi\rangle$, that is, the set of amplitudes $\tilde{a}_j = \langle q_j | \Psi \rangle$, determines all probabilities $Q_j = a_j^2$ for the $q$-measurement outcomes. But the opposite is not true. Taking the set $Q_j = a_j^2$ alone, we lose information stored in the individual phase factors $e^{i\alpha_j}$, and this precludes determining of $|\Psi\rangle$. Even adding probabilities from a different basis will be of little, if any, help. We can write (3) in two representations, e.g.

$$|\Psi(t)\rangle = \sum_j \tilde{a}_j |q_j\rangle = \sum_l \tilde{b}_l |p_l\rangle, \quad \tilde{b}_l = \langle p_l | \Psi \rangle = b_l e^{i(\beta_l - \omega_l t)} \qquad (4)$$

where $p_l$ and $|p_l\rangle$, are, respectively, the eigenvalues and eigenstates of observable $p$ incompatible with $q$. Then the *expanded* set of probabilities $(Q_j, \mathcal{P}_l)$ including all $\mathcal{P}_l = b_l^2$ may at best, involve a huge or even infinite set of calculations to determine either all $\tilde{a}_j$ or all $\tilde{b}_l$ and thus the $|\Psi\rangle$. Let alone the cases when observables $p$ and $q$ form a continuous set.

Also, a single set of probabilities $\mathcal{P}$ may correspond to different sets of $Q$. In a textbook case with a wave packet in a free space we have a fixed spectrum $\mathcal{P}(\omega)$ but different shapes and thereby different sets of $Q(x)$ at different moments of time. This reflects another aspect of the problem – that restricting to probabilities alone leads to loss of information about time evolution of the state, which is stored in the amplitudes.

In contrast, the set of *amplitudes* $\tilde{a}_j$ alone in (3) is sufficient for predicting all $\mathcal{P}_l$ and all $Q_j$. If all $a_j$ embrace the *complete set* of system's observables, they determine probabilities for *any* measurements on this system including the future measurements, so we do not need to "*elicit… one's degrees of belief…*" Throwing away the state $|\Psi\rangle$ together with the respective amplitudes, and restricting exclusively to "*degrees of belief*" converts QM from a working self-consistent theory with predictive power to a cookbook of recipes with an infinite list of "beliefs". It is such a cookbook that was proclaimed a "*Hero's Handbook*" in the title of [4].

{10} "*Just think of a spin $(1/2)$ system. If one has elicited one's degrees of belief for the outcomes of a $\sigma_x$ measurement, and similarly one's degrees of belief for the outcomes of $\sigma_y$ and $\sigma_z$ measurements, then this is the same as specifying a quantum state itself:*" ([1, 3], p. 5, 11)

This might work acceptably only in some special cases, e.g. for a qubit, whose Hilbert space has only two dimensions. But even in this simplest case, it eludes the basic question: where the "*degrees of belief*" come from? They come from the measurements, to begin with! It is a measured quantum system which is the source of "*beliefs*" for all subsequent measurements. And knowing the system's Hamiltonian allows us to learn all about the system just by solving the corresponding equation. Frequently, the Hamiltonian itself is determined from the experimental data (the inverse problem of QM [33]), and then it can be used to reveal the system's states. That answers all relevant questions without any additional interpretations of QM or any cookbooks.

{11} "*For if one knows the quantum state's projections onto three independent axes, then that uniquely determines a Bloch vector, and hence a quantum state. Something similar is true of all quantum systems of all sizes and dimensionality.*" ([2, 4], p. 5, 11)

This is a mix of tautology, ambiguity, and falsehood. The complete set of quantum state's projections determines the state without any references to the Bloch vector. Such reference in this context confuses two fundamentally different concepts. A quantum state is a vector in the respective Hilbert space $\mathcal{H}$, whereas the Bloch vector with polar angle $\theta$ and azimuth $\varphi$ is defined in our physical space $V$ only to represent the system's angular momentum. The dimensionality of $\mathcal{H}$ may be infinite or even uncountable, whereas the Bloch vector lives in 3-D space. Even if we throw away all characteristics other than spin, the accordingly truncated $\mathcal{H}$-space will have dimensionality $D_H = 2s+1$, where $s$ is the spin quantum number. This reduces to $D_H = 3$ for $s =1$ and to $D_H = 2$ for a qubit ($s =1/2$). But even in these simplest cases, the geometrical projections of the Bloch vector are not and cannot be equal to the quantum state's projections, since the latter are generally complex numbers, let alone having a totally different physical meaning. So the Bloch vector is as far from state representation as we can get.

Even if we focus on spin alone, the Bloch vector can represent it only indirectly since spin is not a vector. Its more accurate visual representation is a right circular cone with an opening angle

$$\xi_{s,m} = 2Arc\cos\frac{m}{\sqrt{s(s+1)}} \xrightarrow{m=s} 2Arc\cos\sqrt{\frac{s}{s+1}}, \qquad (5)$$

where *m* is the quantum number of the spin component onto the symmetry axis of the cone [15, 34]. The $\xi_{s,m}$ is minimal at *m* = *s* and can approach zero, thus reducing the corresponding cone to a vector, only at $s \gg 1$. And there is no Bloch vector for a state with $m = 0$.

For $s \geq 3/2$ we may have a set of cones with the common symmetry axis but various $\xi_{s,m}$, all represented by a single Bloch vector (Fig. 1). Already this example gives a visual illustration of the fact that generally there is no one-to-one correspondence between the Bloch vectors and the spin states. Let alone the superpositions of eigenstates $|\xi_{s,m}\rangle$, which may form a continuous set! And let alone all other observables!

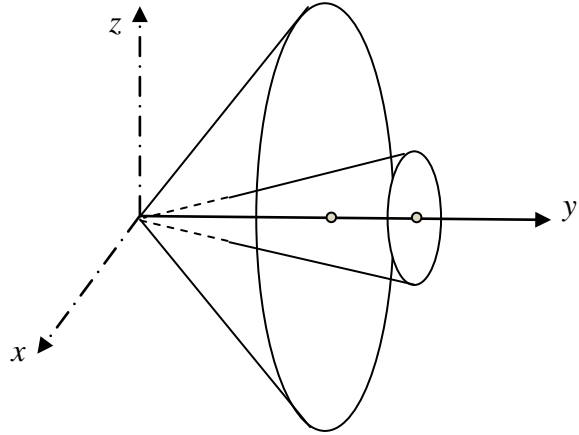

**Fig. 1**
Graphical representation of two spin-eigenstates with $m = s$ and $m = s-1$ along the $y$-axis (not to scale). Both states are represented by the same Bloch vector with $\theta = \varphi = \pi/2$

Consider a few more examples illustrating inadequacies of {10}.
Suppose we have an electron and focus only on its spin, and its initial state is

$$|\Psi\rangle = |\odot\rangle \qquad (6)$$

with the spin pointing toward us along the $x$-axis. Now we are planning a spin measurement along the $z$-axis. The corresponding antipodal points of the Bloch sphere will be the North and South poles, representing the $|\uparrow\rangle$ and $|\downarrow\rangle$ states, respectively. These states are oppositely-directed in $V$ but mutually orthogonal in $\mathcal{H}$. State $|\odot\rangle$ will "collapse" under $S_z$-measurement to either $|\uparrow\rangle$ or $|\downarrow\rangle$ with equal probabilities described by the table

$$\mathcal{P}(s_z) = \frac{1}{2}\begin{pmatrix} 1 \\ 1 \end{pmatrix}, \tag{7}$$

which, according to QBism, should uniquely represent $|\odot\rangle$ in the $s_z$ basis. But it does not since the same table (7) would also represent a *different* initial state, say, $|\rightarrow\rangle$ with the Bloch vector along the *y*-axis. So now we have one table representing two different states. Actual number of states represented by (7) makes a continuous set with all possible Bloch vectors in the (*x*, *y*) plane.

On the other hand, for a massive boson with spin 1 in state (6) with the same direction $|\odot\rangle$ of spin, there will be *three* possible outcomes in the $s_z$-measurements, corresponding to eigenstates with $m_z = 1, 0, -1$. State (6) for such boson is a superposition of these eigenstates, and the corresponding probability table will be

$$\mathcal{P}(s_z) = \frac{1}{4}\begin{pmatrix} 1 \\ 2 \\ 1 \end{pmatrix} \tag{8}$$

So now we have *the same* Bloch vector with different individual tables (7) and (8) for different physical objects. Contrary to claim {10}, there is no one- to-one correspondence between the Bloch vectors and represented systems.

Suppose we have a single photon in a free space. The *complete* set of observables defining its state is, e.g., momentum **p** and polarization. What kind of the Bloch vector would the QBist assign to a non-monochromatic photon state (wave packet) with an indeterminate **p**? What kind of the Bloch vector should be assigned to a *linearly-polarized* monochromatic photon? What kind of the Bloch vector would the QBist associate with a non-stationary state of an excited atom or with an unstable particle in a Gamow state?

{12} "*What this buys interpretatively, beside airtight consistency with the best understanding of probability theory, is that it* (Quantum Bayesianism-M.F.) *gives each quantum state a home. Indeed, a home localized in space and time – namely, the physical site of the agent who assigns it!*" ([2, 4], p. 5, 11)

This is a double-faced argument, with one face suggesting the known solution of the Schrodinger's cat paradox [35, 36]. It states that we can cut and thereby disentangle the potentially infinite von Neumann's chain at the physical site of the agent observing the cat. But this positive aspect of {11} is restricted to states with a single spatially localized site.

The other face shows the argument's flaws in the general case. First, suppose we have an extended wave packet (ideally, a monochromatic wave occupying the whole space), and a human observer as the "*agent*". Where is then the home of the state? "*The physical site of the agent*" is, at the most, his/her Lab. Does this mean that the coherence length of a monochromatic laser beam cannot in principle exceed a few meters, and there are no quantum states outside of my Lab?

As a second example, consider a system of the two separated entangled particles. One is observed by Alice in Los Alamos, and the other by Bob in Brookhaven. If the state's home is "*the physical site of the agent who assigns it* ", then where is the home of such composed state, who assigns it, and what is this assignment?

The third example is a monochromatic light wave from a source S. The wave may be UV for Alice and IR for Bob if they observe it from different spaceships. They can make their measurements when their respective ships closely pass each other, so both will be within the same region of space-time at the moment of measurement. Now the two agents' homes form their temporary common home. Let the first number in a pair $(x, y)$ stand for agent's "*degree of belief*" in UV, and the second one for "*degree of belief*" in IR. Then we have a simple table for their "degrees of belief":

The Truth Table for Alice and Bob

| Object | Agent | Home | Measured state | Degrees of belief |
|---|---|---|---|---|
| Plane monochromatic light wave | Alice | Alice's site (spaceship passing the star Umbra) | UV | (1, 0) |
| | Bob | Bob's site (another spaceship passing Umbra at the same time) | IR | (0, 1) |

How does it fit into the Bayesian scheme? Here we have one "*home*" with two equally qualified agents in it. But their "*degrees of belief*", even when objectified (derived from their respective *measurements*), totally disagree with one another even though referring to *the same system*. And yet their mutually opposing beliefs are both true because the observed frequency is associated not only with the wave itself but also with a RF from which it is measured. In this example, both agents are just recording what *is* determined by measured state (pure ensemble of photons) *and* experimental conditions.

In another version, Bob can fly away for vacation and Alice stay in her ship, now stationary relative to S, but with her measuring device on the edge of a rapidly rotating optical table. The device records interchangeably UV and IR for the same radiation. Now we have a permanent site with one agent changing her degrees of belief every fraction of a second.

In both cases, state and the conditions of its measurement are primary, not "*degrees of belief*" derived from measurements.

{13} "*If there is a right quantum state, then why not be done with all this squabbling and call the state a physical fact to begin with? It is surely external to the agent if the agent can be wrong about it. But, once you admit that (and you should admit it), you're* *sunk: For, now what recourse do you have to declare no action at a distance when a delocalized quantum state changes instantaneously?*

*Here I am with a physical system right in front of me, and though my probabilities for the outcomes of measurements I can do on it might have been adequate a moment ago, there is an objectively better way to gamble now because of something that happened far in the distance?*(*Far in the distance and just now.*) *How could that not be the signature of action at a distance?*" ([2, 4], p. 5, 12)

"*Without the protection of truly personal quantum state assignments, action at a distance is there as doggedly as it ever was.*" ([2, 4], p. 6, 12)

The delocalized quantum states are well studied and there is an extensive literature on them [12, 18, 36-44]. Both – the theory considering quantum states as objective reality, and experimental evidence – show no action at a distance. Consider again an entangled state of two widely separated particles A and B. The state of B does change instantaneously at the moment of measurement on A. But this kind of change is not due to any instantaneous signaling (the latter would be equivalent to "*action at a distance*" in Bayesian terminology), because of the pre-existing non-zero probability for B to be found in the new state even in the absence of any measurements on A. In other words, the corresponding changes of states in A and B, albeit correlated, are not in the cause and effect relationship (see, e.g., [45], Sec. 11.10). In this case the relativity of simultaneity is not a problem either. If the observers Alice and Bob perform the corresponding measurements on their respective particles A and B, and the interval AB is space-like, we can always find two different RF, K and K', such that chronology of measurements is (A, B) in K and (B, A) in K'. This does not constitute any contradictions because there is no information flow between A and B. It is the essence of quantum non-locality – the correlations between separated events without any information exchange between them (a detailed discussion can be found, e.g., in [12, 14]).

A single particle in a state with indeterminate coordinate (e.g., a wave packet in a free space) is a different case of quantum non-locality (albeit rarely emphasized as such). But here, too, there is no superluminal energy or information transfer between different parts of the packet in the process of its collapse to some definite location [12].

### 3. Space-time picture of quantum collapse

{14} "*Take, for instance, the infamous "collapse of the wave function," wherein the quantum system inexplicably transitions from multiple simultaneous states to a single actuality.*" [1]

Here we see again miserable consequences of using the term "collapse" instead of *quantum jump* or *instant reconfiguration* (*IR*) (not to be confused with IR!)). The "*infamous collapse of the wave function*" totally ignores the opposite side of the same event – the accompanying "explosion" in another basis [15]. In that basis, *the same resulting state* is again a superposition of multiple simultaneous states. An extended wave packet collapsing to a sharp spike in configuration space largely expands in the momentum space. These are just different, but equally legitimate "faces" of the same final state. But promoting the final state in {14} to the status of *"a single actuality"* precludes admitting that it is a superposition of states in another basis, and this flatly contradicts QM. Or else, the *initial* superposition of "*multiple simultaneous states*" must have the same status of "*a single actuality*" as well, for that initial state, being "*multiple simultaneous states*" in configuration space, may be a sharp spike in the momentum space. But admitting this would now contradict {14}. Article [1] shows only one face of the process as a single possibility, which contributes to misleading statements like {14}. Thus, {14} either contradicts itself, or clashes with well-established principles of QM.

{15} "*A quantum particle can be in a range of possible states. When an observer makes a measurement, she instantaneously "collapses" the wave function into one possible state. QBism argues that this collapse isn't mysterious. It just reflects the updated knowledge of the observer. She didn't know where the particle was before the measurement. Now she does.*" [1]

This statement perpetuates the same fatal misconception as the previous one. "*One possible state*" in configuration space is a broad range of possible states in the momentum space. The same observer who ".. *didn't know where the particle was before the measurement* " well knew (or could have calculated from the initial state) particle's "location" or at least, range of locations in the momentum space. Either knowledge was part of the total information available to her. So should or shouldn't she promote the initial situation to a "*single actuality*"? The same question arises if we turn to the post-measurement situation. Now the observer's knowledge of particle's initial location in the momentum space is irretrievably lost with the *IR* of particle's state under measurement. So should or shouldn't the observer promote this new situation to a "*single actuality*"?
  QBism as represented in [2-4] does not define mechanism of formation of "*degrees of belief*" of an agent. This provokes a few questions to a QBist.
  1) Consider position measurement of an electron described to a high accuracy by de Broglie's wave with definite energy. The agent has no preliminary information about this state. What is numerical measure of his/her degree of belief to find the electron within a specific detector?  Under given conditions the only way to compile any "*list of assignments*" is performing a huge set of measurements whose outcomes are determined by the preexisting state, not vice versa. Will QBism still claim that the obtained "*degrees of belief*" define the state?
 2) If "Yes", then which state do they define? The one existing before the measurements, that is, before formation of any "*degrees of believe*" and creating the "*list of assignments*", or the new one created by measurements?
 3) Suppose we had not been informed about the energy of the coming electron, so we do not have any "*degrees of belief*" for energies. Will the QBist say that "*what is*" there has no such attribute as energy? Next, does the electron have a certain but just unknown to us position at any moment before position measurement? If yes, then Bayesian comes up with paradoxical statement that instead of having definite energy and indefinite position, the coming electron has at each moment definite position and no such thing as energy. If not, will the Bayesian admit that position indeterminacy is an objective characteristic of the coming electron?
  4) Suppose we have a periodic array of equidistant identical detectors, and after a series of trials, all of them showed equal probability of capturing an electron. According to conventional QM, this may be a signature of at least two different states: a monochromatic plane wave or a standing wave with a wavelength equal to distance between the neighboring detectors. Which one out of these different states would the QBist associate with his/her degrees of belief?

It is easy to distinguish between the two possibilities in the last question, e.g., by randomizing positions of detectors. If the local probabilities remain constant and insensitive to location, we have the first possibility; otherwise – the second one. But if the observer is negligent or incompetent or both, the two objectively different physical states will be represented as one in his/her mind. Thus, the QBist's interpretation eliminates the one- to-one correspondence between the agent's vision and reality, which is absolutely necessary for any progress in our understanding of the world. Ignoring this point makes the QBism questionable already on the conceptual level.

Now let us analyze in more detail the possible experiment for position measurement on a state with definite momentum. We place the set of identical detectors in the electron's way, so there is an equal chance for the electron to be captured within an arbitrarily located volume element $\Delta V$. Suppose one of detectors fired, and denote the corresponding coordinate and moment of time as (0, 0). We know immediately that the pre-existing wave packet has "collapsed" to the origin of our inertial frame K. In total contrast with Bayesian statements, this "change of beliefs" in our consciousness happens only because of real changes out there – the firing of detector at (0, 0) with all the others remaining idle. According to QM, the "collapse" of the probability cloud to the "selected" detector happened instantaneously. The probabilities became zero instantly everywhere except for the one at (0, 0), which jumped to 1. And this is not associated with any signaling: there was no probability flow converging to the origin. The cloud just instantly disappeared everywhere outside the small cell at (0, 0), and reemerged as a spike with probability 1 within it. This *IR* of the probability cloud has a conservation law as its primary cause, rather than re-evaluation of probabilities by an observer, which is merely a secondary action. It is just one of manifestations of quantum non-locality, a mundane phenomenon in the quantum world, but something totally exotic in classical world.

All this occurs without conflicting with SR because the entity doing it – the *probability distribution* $\mathcal{P}(\mathbf{r})$ – is only indirectly connected with anything physical. But the connection (and a very clever one!) exists, which is evident, e.g., from the fact that the product $q_e \mathcal{P}(\mathbf{r})$ determines the charge density distribution and hence the current and the corresponding magnetic field produced, e.g., by an atomic electron with nonzero angular momentum *L*. This field can, to a high accuracy, be treated classically for a continuously evolving state described by a wave equation, but undergoes the *IR* together with $\mathcal{P}(\mathbf{r})$ in a position measurement. All this shows that the set of probabilities is an objective characteristic of an objective state rather than "*degrees of belief*"of some observer. By all evidence available today, Nature did not need any observers in order to operate by Her own rules. It is the observer who needs to develop the ability to grasp the intricacies of Nature for their consistent description.

There still remains a great mystery though – how can Nature, operating by Her rules, develop Her small part into self-conscious intelligent observers perceiving and trying to understand Nature? This is a big question going far beyond today's physics. I only pose it here because it is related to the discussed topic.

## 4. The Wigner's friend paradox

The next important issue discussed in QBism is the "Wigner's friend" paradox. It is the story of two "*agents*", Wigner and his friend, both studying a certain quantum system:

{16} "*Suppose they both agree that some quantum state $|\Psi\rangle$ captures their mutual beliefs about the quantum system.* (But according to [2, 3], quantum systems do not exist! - M. F.) *Furthermore suppose they agree that at a specified time the friend will make a measurement on the system of some observable (outcomes i = 1, . . . , d). Finally, they both note that if the friend gets outcome j, he will (and should) update his beliefs about the system to some new quantum state $|j\rangle$. There the conversation ends and the action begins: Wigner walks away and turns his back to his friend and the supposed measurement. Time passes to some point beyond when the measurement should have taken place.*
*What now is the "correct" quantum state each agent should have assigned to the quantum system? We have already concurred that the friend will and should assign some $|j\rangle$. But what of Wigner?*" ([2, 4], p. 6, 14)

In conventional QM, there is no paradox here. Actually, the "Wigner's friend" is an extended version of the "Schrodinger's cat" with a distinction that instead of the cat with two eigenstates ($|\text{alive}\rangle$ or $|\text{dead}\rangle$) we have an object with an arbitrary number *d* of eigenstates $|i\rangle$. Once it is given that the object before the measurement has a state $|\Psi\rangle$ of its own, we can write this state as

$$|\Psi\rangle = \sum_{i=1}^{d} \tilde{c}_i |i\rangle \quad , \tag{9}$$

where $\tilde{c}_i = \langle i|\Psi\rangle$ are the corresponding amplitudes. During the measurement, the studied object gets entangled with the apparatus and, in the given context, with the friend; finally, the entangled superposition collapses to a single product state No. *j*, which is the measurement outcome. The whole process can be written as

$$|\Psi\rangle = \sum_{i=1}^{d} \tilde{c}_i |i\rangle \quad \Rightarrow \quad \sum_{i=1}^{d} \tilde{c}_i |i\rangle \otimes |A_i\rangle \otimes |F_i\rangle \quad \Rightarrow |j\rangle \otimes |A_j\rangle \otimes |F_j\rangle \tag{10}$$

$$t < 0 \qquad\qquad 0 < t < t_D \qquad\qquad t > t_D$$

Here $|A_i\rangle$ is the *i*-th pointer state of the apparatus, $|F_i\rangle$ is the state of friend's awareness in the outcome $|i\rangle$, and $t_D$ is generally unknown but usually very small decoherence time. The first two expressions describe the continuous evolution of object's state $|\Psi\rangle$ and then of combined system (object + apparatus + Wigner's friend). The transition to the third expression is a discontinuous jump completing the measurement. After the moment $t_D$, the difference between psychological states of Wigner and his friend can be represented by the respective column matrices

$$\text{Wigner:} \begin{pmatrix} \mathcal{P}_1 \\ ... \\ \mathcal{P}_i \\ ... \\ \mathcal{P}_d \end{pmatrix} ; \qquad \text{Wigner's friend:} \begin{pmatrix} 0 \\ ... \\ 1 \\ ... \\ 0 \end{pmatrix} = |F_j\rangle \qquad (11)$$

Here $\mathcal{P}_i \equiv |\tilde{c}_i|^2 = c_i^2$ is probability of (or "*degree of belief*" in) the outcome $|i\rangle$. The column on the right is the matrix representation of state $|F_j\rangle$.

According to QBist's own definition (quote {11}), the true quantum state is the one determined "at home" – "*...a home localized in space and time—namely, the physical site of the agent who assigns it!*". This resolves the "paradox" – the true state after the measurement is $|j\rangle$, which is also encoded in mental state $|F_j\rangle$ of Wigner's friend.

What we read next, initially seems to follow the same track:

{17} "*If he were to consistently dip into his mesh of beliefs, he would very likely treat his friend as a quantum system like any other: one with some initial quantum state $\rho$ capturing his (Wigner's) beliefs of it (the friend), along with a linear evolution operator U to adjust those beliefs with the flow of time. Suppose this quantum state includes Wigner's beliefs about everything he assesses to be interacting with his friend — in old parlance, suppose Wigner treats his friend as an isolated system. From this perspective, before any further interaction between himself and the friend or the other system, the quantum state Wigner would assign for the two together would be $U(\rho \otimes |\Psi\rangle\langle\Psi|)U^+$ — most generally an entangled quantum state. The state of the system itself for Wigner would be gotten from this larger state by a partial trace operation; in any case, it will not be an $|j\rangle$.*" ([2, 4], pp. 6-7, 14)

This part may be correct, but only as far as it reflects the history *before* $t_D$. On that stage the expression

$$|\tilde{\Phi}\rangle \equiv U(\rho \otimes |\Psi\rangle\langle\Psi|)U^+, \qquad (12)$$

if treated properly, must (for $0 < t < t_D$) be equivalent to the entangled superposition ("*...an entangled quantum state*") in the middle part of (10), and as time passes, it must approach the jump towards the right part. But the next argument stops short of it:

{18} "*Wigner holds two thoughts in his head: 1) that his friend interacted with a quantum system, eliciting some consequence of the interaction for himself, and 2) after the specified time, for any of Wigner's own further interactions with his friend or system or both, he ought to gamble upon their consequences according to $U(\rho \otimes |\Psi\rangle\langle\Psi|)U^+$.*

*One statement refers to the friend's potential experiences, and one refers to Wigner's own.*
*So long as it is kept clear that $U(\rho \otimes |\Psi\rangle\langle\Psi|)U^+$ refers to the latter—how Wigner should gamble upon the things that might happen to him—making no statement whatsoever about the former, there is no conflict."* ([2, 4] p. 7, 14)

Quite the contrary – now *there is* a conflict! Before the moment $t_D$, the true state was described by (12). After that moment, (12) becomes outdated. According to conventional QM (and to Bayesian own definition of the "home state"!), the true physical state of the studied object *after the measurement* ($t > t_D$) is $|j\rangle$. Wigner's friend knows this with certainty just by taking a look at his apparatus. But Wigner, being "not at home", still assigns to it only a "degree of belief" $\mathcal{P}_j$ (which is generally a small fraction of $\mathcal{P}=1$).

Why keep today the yesterday's information about something rapidly changing? Even if all communications are blocked, it is not a good excuse to linger in the middle of Eq. (10). It would be right to say that the current state of the system is just unknown to an outsider – quite a natural thing in both – classical and quantum physics! To make it known, it is necessary to look at the apparatus' pointer or ask the one who had done it. One cannot express this better than Pauli [44]

{19} "*…personal qualities of the observer do not come into the theory in any way—the observation can be made by objective registering apparatus, the results of which are objectively available for anyone's inspection*" (quoted in [2, 4], p. 7, 14).

Instead, the QBist decides that Wigner or any other outsider must stick with outdated information given by the left column in expression (11), that is, remain in the middle of (10). This contradicts his/her own definition of true quantum state as the "home state" of the observer, albeit this is also debatable as shown above in an example with two spaceships. The QBist "eliminates" such contradiction by marginalizing the fundamental concept of quantum states [3], or even proclaiming them nonexistent [2]. But this creates a new contradiction: what does "home state" mean if "*quantum states do not exist*"? Why at all should we "*give home*" to something nonexistent?

We see that promoting personal "*degrees of belief*" produces the opposite of what is claimed – it confuses the discussed issues and makes the whole theory inconsistent.

### 5. Measurement and indeterminacy

{20} "*QBism says when an agent reaches out and touches a quantum system—when he performs a quantum measurement—that process gives rise to birth in a nearly literal sense. … It is the "outcome," the unpredictable consequence for the very agent who took the action. John Archibald Wheeler said it this way, and we follow suit, "Each elementary quantum phenomenon is an elementary act of 'fact creation.'"* ([29], p. 8)

This is true, but it is important to emphasize that in the discussed situation a newly created fact is *not* totally unpredictable. It is of the kind allowed by Nature, rather than some miracle from a fairy tale. Second, it was just lying there dormant as one of the preexisting potentialities waiting for actualization. And third, each actualization had an *exactly defined probability* determined by the pre-existing state of the system and the environment.

{21} *"All that matters for a personalist Bayesian is that there is uncertainty for whatever reason. There might be uncertainty because there is ignorance of a true state of affairs, but there might be uncertainty because the world itself does not yet know what it will give –i.e., there is an objective indeterminism."* ([2], footnote 14, p. 8)

This is 100% true.

{22} *"… QBism finds its happiest spot in an unflinching combination of "subjective probability" with "objective indeterminism.""* ([2], footnote 14, p. 8)

It is unclear how the two may form an "*unflinching combination*". This is an illustration of *subjective indeterminism* of terminology used by some QBists – kicking the words back and forth between different meanings. If quantum states with their objective characteristics do not exist as stated in Sec. 2 of [2], than *"objective indeterminism*" of what is meant here? There is no way for *truly objective* indeterminism to coexist peacefully with *subjective* probabilities.

{23} " *The notorious "collapse of the wave-function" is nothing but the updating of an agent's state assignment on the basis of her experience.*" ([3], p. 3)

As already emphasized above, "*the updating*" is just recording an actual event, e.g., firing of the respective detector.

{24} "*… contrary to the view of some physicists and many philosophers of science, there is no clash between quantum mechanics and special relativity*." ([3], p. 5)

True.

{25} " *No agent can move faster than light: the space-time trajectory of any agent is necessarily timelike. Her personal experience takes place along that trajectory. Therefore when any agent uses quantum mechanics to calculate "[cor]relations between the manifold aspects of [her] experience", those experiences cannot be space-like separated. Quantum correlations, by their very nature, refer only to time-like separated events: the acquisition of experiences by any single agent. Quantum mechanics, in the QBist interpretation, cannot assign correlations, spooky or otherwise, to space-like separated events, since they cannot be experienced by any single agent. Quantum mechanics is thus explicitly local in the QBist interpretation*." ([3], p. 6)

According to the real QM, *quite the opposite is true*: QM does assign correlations to a broad class of space-like separated events (see, e.g., [12, 15, 35-45]). Moreover, most of the observable Universe is described by the set of data available to any single agent from information brought by light or subluminal particles. Therefore the same QBist logic "proves" that all observable Universe "*is explicitly local*."

The falsehood of {25} is seen from a few very simple examples. First, our experience originates not only along our personal trajectory, but as just mentioned, from innumerable carriers of the outside information. In many cases, the respective messages come from different, sometimes opposite directions from space-like separated sources. Their signals make a huge contribution to our personal experience about the world at one moment of our time.

Consider the observers Alice and Bob, and Celia midway between them. Celia sends to both a pair (AB) of spin-entangled particles with the zero net spin – A to Alice and B to Bob. The paths of both particles are time-like, but the arrivals to their respective destinations make the space-like interval. Following the preliminary protocol, Alice and Bob simultaneously measure spins of the arrived particles in the same basis. Celia does not know the individual outcomes, but being an expert in QM, she knows in advance that if A is found, say, in a state $|\uparrow\rangle_A$, then B will collapse to state $|\downarrow\rangle_B$, and vice versa. In contrast to statement {25}, Celia *can* assign correlations to such space-like separated events. The reports from the experimenters only confirm her predictions, and even tell her the exact individual results. In this way, QM shows its *nonlocal* aspect to a local observer. (See, e.g., [34])

{26} "*An agent's assignment of probability* 1 *to an event expresses that agent's personal belief that the event is certain to happen. It does not imply the existence of an objective mechanism that brings about the event. Even probability* 1 *judgments are judgments. They are judgments in which the judging agent is supremely confident.*" ([3], p. 6)

The pre-modern beliefs in flat immovable Earth at the center of the world were all "*probability* 1 *judgments*" of all agents, including the most informed intellectuals of the time. Later, the expanding set of data collected by such agents as Copernicus, Tycho Brahe, Galileo, showed a different, increasingly complex, picture of the world. If, according to QBism, the new picture does not imply the existence of objective reality behind it but is nothing more than the changing collective judgments, then why those judgments have been changing so radically? The only rational answer is: Because of the new information obtained by the new, better equipped agents. But that would sound rational only if we get an equally reasonable answer to: *Information about what*?

{27} "*But when we attempted to understand phenomena at scales not directly accessible to our senses, our ingrained practice of divorcing the objects of our investigations from the subjective experiences they induce in us got us into trouble. While our efforts at dealing with phenomena at these new scales were spectacularly successful, we have just as spectacularly failed for almost a century to reach any agreement about the nature or meaning of that success.*"

All disagreements are natural and even necessary in any real science. Since most of them are eventually settled by factual experimental evidence, only the remaining ones are of the conceptual nature about interpretation. And QM is not the only example. Any area of physics emerging after significant breakup with the previous concepts is vibrant with controversies. SR is up to this day simmering with debates (see, e.g., [46-57]) about simultaneity, relativistic causality, superluminal signaling, relativistic invariance, length contraction and time dilation, mass-energy equivalence, relativistic mass, and so on. Such debates are natural when new evidence surpasses the previous intuition based on the ingrained concepts of the absolute time and absolute mass.

Today, numerous interpretative disagreements continue simmering despite the astounding successes of conventional QM in description of observed phenomena and prediction of new effects. And the controversies result from unavoidable difficulties associated with necessity to go beyond the classical concepts and to grasp the newly discovered dimensions of reality.

## 6. Conclusions

The mathematical structure of QM is self-consistent, and experiments confirm all the discussed effects. On the other hand, the effects themselves and the basic concepts behind them are to a high degree alien to our classical intuition. There are two possible ways to deal with the situation: either to enrich our imagination and intuition by accepting newly discovered dimensions of reality and getting used to them, or to try to describe them in terms of the old concepts. But as all history of science shows, the latter is a blind venue.

QBism declares the quantum states subjective, instead of admitting that classical intuition alone is inadequate to embrace the intrinsically probabilistic nature of the quantum world. As Nick Herbert put it [58]:

"*QBism fails as an interpretation because it does not even address the question of what is the nature of a world that gives rise to these particular probabilities… It says nothing about what this world does. Simply put, QBism is not an interpretation of the quantum world*."

The Qbism's high assertion quoted in the epigraph is degraded to zero by focusing on the observer alone and downplaying the world around.

We are forced to conclude that the Bayesian interpretation of QM is faulted and most of its claims are not substantiated.

## Acknowledgments

I am grateful to Nick Herbert for turning my attention to the discussed topic and for his valuable comments.

## References

1. Katherine Taylor, A Private View of Quantum Reality,
   *Quanta Magazine*, 08/17/2015;
   Also Amanda Gefter, A private view of quantum reality, <u>*Science*</u>, 06/14/2015, or
   https://www.wired.com/2015/06/private-view-quantum-reality)
2. Christopher A. Fuchs, QBism, the Perimeter of Quantum Bayesianism,
   arXiv:1003.5209v1 [quant-ph] 2010
3. C.A. Fuchs, N. D. Mermin, R. Schack,
   An introduction to QBism with an application to the locality of quantum mechanics,
   arXiv:1311.5253 (2013)
4. Christopher A. Fuchs, Blake C. Stacey,
   QBist Quantum Mechanics: Quantum Theory as a Hero's Handbook,

   arXiv:1612.07308v1 [quant-ph]
5. Mermin N. D., Why QBism is not the Copenhagen interpretation and what John Bell might have thought of it,   arXiv:1409.2454 [quant-ph]
6. L. Marchildon, Why I am not a QBist, *Found. of Phys.*, **45** (7), 754-761, 2015
7. M. Nauenberg, QBism and locality in quantum mechanics,
    *Amer. Journ. of Phys.* **83**, 197 (2015)
8. H. C. von Baeyer, *QBism: The Future of Quantum Physics*,
    Harvard University Press, 2016
9. A. C. Elitzur and L. Vaidman, Quantum-Mechanical Interaction-Free Measurements, Foundations of Physics, **23**, 7, 987-997, (1993)
10. P. G. Kwiat, H. Weinfurter, T. Herzog, A. Zeilinger, and M. A. Kazevich, Interaction-Free Measurement, Phys. Rev. Lett., **74**, 24, 4763 – 66, (1995)
11. Y. Aharonov and D. Bohm,
    Significance of Electromagnetic Potentials in the Quantum Theory,
    *Phys. Rev*. **115**, 485 (1959)
12. M. Fayngold, How the instant collapse of a spatially-extended quantum state is consistent with relativity of simultaneity, *Eur. Journ. Phys*. **37** (6) (2016); arXiv:1605.05242 [physics.gen-ph]
13. J. von Neumann, *Mathematical Foundations of Quantum Mechanics*, Princeton Univ. Press, 1955
14. J. S. Bell, "Against 'Measurement'," *Phys. World* **3**, (8) 33 (1990)
15. M. Fayngold, V. Fayngold, *Quantum Mechanics and Quantum Information*, Wiley-VCH, Weinheim, 2013
16. J. B. Hartle, "Quantum Mechanics of Individual Systems,"
    *Am. J. Phys*. **36**, 704 (1968)
17. D. I. Blokhintsev, *Principles of Quantum Mechanics*, Allyn & Bacon, Boston, 1964
18. W. H. Zurek, Decoherence, einselection, and the quantum origins of the classical, arXiv:quant-ph/0105127v3 Jun 2003
19. R. W. Spekkens, Evidence for the epistemic view of quantum states: A toy theory *Phys. Rev. A*, **75**, 032110 (2007)
20. J. S. Bell, On the problem of hidden variables in QM,
    *Rev. Mod. Phys*., **38**, 447 (1966)
21. N. Herbert, *Quantum Reality*, Doubleday, New York, 1987
22. J. F. Clauser, M. A. Horne, A. Shimony, and R. A. Holt,
    Proposed experiment to test local hidden-variable theories,
    *Phys. Rev. Lett*., **23**, 880-884 (1969)
23. J. F. Clauser and A. Shimony, Bell's theorem: experimental tests and implications, *Rep. Prog. Phys*., **41**, 1881 (1978)
24. A. Aspect, P. Grangier, and G. Roger,
    Experimental realization of EPR-Bohm Gedanken experiment: a new violation of Bell's inequalities, *Phys. Rev. Lett*., **49**, 91-94 (1982)
25. A. Aspect, J. Dalibard, and G. Roger,
    Experimental test of Bell's inequalities using time-varying analyzers,
    *Phys. Rev. Lett*., **49**, 1804 (1982)
26. B. de Finetti, *Theory of Probability*, Wiley, 1990
27. Robert F. Nau, "De Finetti was Right: Probability Does Not Exist",

*Theory and Decision*, **51**, (2-4), 89–124 (2001)
28. Frank A. Haight, *Handbook of the Poisson Distribution*,
    John Wiley & Sons, New York, 1967
29. S. Saunders, What is probability? in *Quo Vadis Quantum Mechanics*?
     (Springer) 2005, p. 209;  arXiv:quant-ph/0409144
30. I. J.Good, "46656 varieties of bayesians", in *Good Thinking*:
    *The Foundations of Probability and Its Applications* (University of Minnesota Press)
     1983, p. 20.
31. E. T. Jaynes, *Probability Theory: The Logic of Science*,
    Cambridge University Press, 2003.
32. Wigner E. P. The probability of the existence of a self-reproducing unit,
     in *The Logic of Personal Knowledge: Essays Presented to Michael Polanyi on his
     Seventieth Birthday*  (Routledge & Kegan Paul, London) 1961, pp. 231
33. B. N. Zakhariev and V. M. Chabanov, *Submissive Quantum Mechanics: New
    Status of the Theory in Inverse Problem Approach*,
    Nova Science Publishers Inc., New York, 2007
34. M. Fayngold, Multi-Faced Entanglement, arXiv:1901.00374 [quant-ph]
35. J. S. Bell, On the EPR paradox, *Physics*, **1**, 195 – 200 (1964)
36. P. H. Eberhard, Bell's theorem and the different concepts of locality,
     *Nuovo Cim.*, **46**, 392 (1978)
37. N. Herbert, FLASH: a superluminal communicator based upon a new kind of
    quantum measurement, *Found. Phys.*, **12**, 1171 (1982)
38. W. K. Wootters and W. H. Zurek, A single quantum cannot be cloned,
    *Nature*, **299**, 802 (1982)
39. D. Dieks, Communication by EPR devices, *Phys. Lett*. A, **92** (6), 271 (1982)
40. S. J. van Enk, No-cloning and superluminal signaling,
    arXiv:quant-ph/9803030 (March 1988)
41. P. H. Eberhard and R. R. Ross, Quantum field theory cannot provide faster
    than light communication, *Found. Phys. Lett.*, **2**, 127 – 149  (1989)
42. F. De Martini, F. Sciarrino, and C. Vitelli,
    Entanglement and non-locality in a micro-macroscopic system,
    arXiv:0804.0341 (2008)
43. A. Aspect, To be or not to be local, *Nature*, **446**, 866 – 867 (2007)
44. W. Pauli, *Writings on Physics and Philosophy*,
    Ed. by C. P. Enz and K. von Meyenn, (Springer-Verlag, Berlin, 1994)
45. M. Fayngold, *Special Relativity and How It Works*, Wiley-VCH, Weinheim,
    Sec. 11.10, 2008
46. O. M. Bilaniuk, V. K. Deshpande, E. C. G. Sudarshan,  Meta-relativity,
    *Am. J. Phys.*, **30**, 718-723 (1962)
47. G. Feinberg, Possibility of Faster-than-Light Particles,
    *Phys. Rev.*, **159**, 1089-1105  (1967)
48. P. L. Csonka, Causality and Faster than Light Particles,
     *Nucl. Phys.*, **B21**, 436-444   (1970)
49. R. Fox, C. G. Kuper, S. G. Lipson, Faster-than-light group velocities and causality
     violation, *Proc. Roy. Soc. London*, **A316**, 515-524 (1970)
50. G. A. Benford,  D. L. Book, W. A. Newcomb, The Tachyonic Antitelephone,

    *Phys. Rev. D.*, **2** (2), 263 (1970)
51. E. Recami, Classical Tachyons and Possible Applications,
    *Rivista Nuovo Cim.* **9** (6), 1-178 (1986)
52. M. Ya. Azbel', Superluminal velocity, tunneling traversal time and causality,
    *Solid State Commun.*, **91** (6), 439–441 (1994)
53. P. C. Peters, Does a group velocity larger than *c* violate relativity?
    *Am. J. Phys*. **56** (2), 29-131 (1988)
54. T. Sandin, In Defense of Relativistic Mass, *Am. J. Phys.*, **59** (11), 1032 – 1036 (1991)
55. G. Oas, On the abuse and use of relativistic mass, arXiv: *Physics*/0504110v2 (2005)
56. L. B. Okun, The Virus of Relativistic Mass in the Year of Physics.
    In the "*Gribov Memorial Volume*: *Quarks*, *Hadrons*, *and Strong Interactions*",
    World Scientific, Singapore (2006)
57. M. Fayngold, Three +1 Faces of Invariance, arXiv:1001.0088 [physics.gen-ph]
58. N. Herbert, Private communication (2020)